\documentclass{goeproc}

\pdfoutput=1

\def\cobold{CO$^{\sf 5}$\-BOLD}

\begin{document}

\title{Recent progresses in the simulation of small-scale magnetic fields}

\author[ ]{O.~Steiner}

\affil[ ]{Kiepenheuer-Institut f\"ur Sonnenphysik, Freiburg, Germany}
\affil[ ]{\textit{Email:} steiner@kis.uni-freiburg.de}

\runningtitle{Simulation of small-scale magnetic fields}
\runningauthor{O.~Steiner}

\firstpage{321}

\maketitle

\begin{abstract}
New high-resolution observations reveal that small-scale magnetic flux 
concentrations have a delicate substructure on a spatial scale of $0.1''$. Its 
basic structure can be interpreted in terms of a magnetic flux sheet or tube that  
vertically extends through the ambient weak-field or field-free atmosphere
with which it is in mechanical equilibrium. A more refined interpretation 
comes from new three-dimensional magnetohydrodynamic simulations 
that are capable of reproducing the corrugated shape of magnetic flux 
concentrations and their signature in the visible continuum. Furthermore it is 
shown that the characteristic asymmetric shape of the contrast profile of facular 
granules is an effect of radiative transfer across the rarefied atmosphere of the 
magnetic flux concentration. I also discuss three-dimensional radiation magnetohydrodynamic 
simulations of the integral layers from the top of the convection zone to the 
mid-chromosphere. They show a highly dynamic chromospheric magnetic field,
marked by rapidly moving filaments of stronger than average
magnetic field  that form in the compression zone downstream and along propagating
shock fronts. The simulations confirm the picture of flux concentrations that strongly 
expand through the photosphere into a more homogeneous, space filling chromospheric 
field. Future directions in the simulation of small-scale magnetic fields are
indicated by a few examples of very recent work.
\end{abstract}

\section{Introduction}
\label{sect_int}

With ``realistic simulations'' computational physicists aim  at imitating 
real physical processes that occur in nature. In the course of rebuilding 
nature in the computer, they aspire to a deeper understanding of the 
process under investigation. In some sense the opposite approach is taken
by computational physicists that aim at separating the fundamental
physical processes by abstraction from the particulars for obtaining
``ideal simulations'' or an analytical model of the essential physical process. 
Both strategies are needed and are complementary as can be seen for example in
Section \ref{sect_fac} on the physics of faculae. In this paper,
however, we mainly focus on ``realistic simulations'' and comparison
with observations.

The term small-scale flux concentration is used here to designate the 
magnetic field that appears in G-band filtergrams as bright tiny objects 
within and at vortices of intergranular lanes.  They are also visible in 
the continuum, where they are called \emph{facular points} \citep{mehltretter74},
while the structure made up of bright elements is known as the 
\emph{filigree} \citep{dunn+zirker73}. In more recent times, the 
small-scale magnetic field was mostly observed in the G band 
(a technique originally introduced by \citealp{muller1985})
because the molecular band-head of CH that constitutes the G band 
acts as a leverage for the intensity contrast 
\citep{rutten99,rutten+al01,sanchez-almeida+al01,shelyag+al04,steiner+al01}.
Being located in the blue part of the visible spectrum, this 
choice also helps improving the diffraction limited spatial resolution 
and the contrast in the continuum. 

Small-scale magnetic flux concentrations are studied for
several reasons:
\begin{list}{-}{\setlength{\topsep}{4pt}\setlength{\parsep}{0.0mm}
                      \setlength{\itemsep}{4pt}\setlength{\leftmargin}{10pt}}
\item[-] Since they make up the small end of a hierarchy of magnetic structures 
         on the solar surface, the question arises whether they
         are ``elemental'' or whether yet smaller flux elements exist.
         How do they form? Are they a surface phenomenon?
         What is their origin?
\item[-] Near the solar limb they can be identified with faculae, known to
         critically contribute to the solar irradiance variation.
\item[-] They probably play a vital 
         role in the transport of mechanical energy to the outer 
         atmosphere, e.g., by guiding and converting magnetoacoustic 
         waves generated by the convective motion and granular buffeting.
\end{list}


\section{The basic structure of small-scale magnetic flux concentrations}
\label{sect_obs}

\begin{figure}
\centering

\begin{minipage}[b]{0.475\textwidth}
\includegraphics[width=1.0\linewidth]{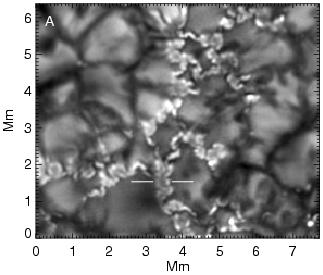}
\end{minipage}\hfill
\begin{minipage}[b]{0.475\textwidth}
\includegraphics[width=1.0\textwidth]{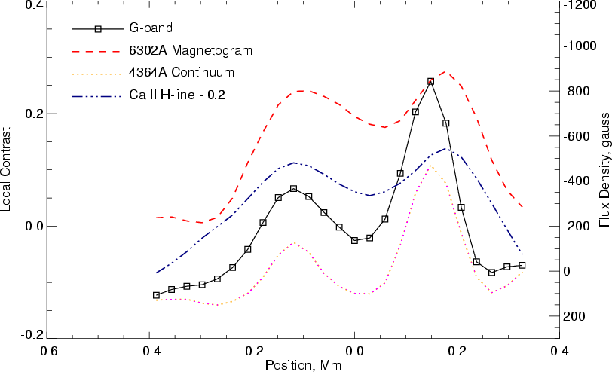}
\vspace*{0.08\linewidth}
\end{minipage}
\caption{Left: G-band filtergram showing the ribbon-like shape
of magnetic flux concentrations. Right: G-band intensity (solid curve) 
along the section indicated by the horizontal white lines in
the image to the left. Also shown are the magnetogram signal, the
continuum intensity at 436.4~nm, and the Ca II H-line intensity 
down-shifted by 0.2. From \citet{berger+al04}.
\label{steiner_fig01}
}
\end{figure}

Recent observations of unprecedented spatial resolution with the 
1~m Swedish Solar Telescope by \citet{berger+al04} and \cite{vandervoort05}
reveal  G-band brightenings in an active region as delicate, corrugated 
ribbons  that show structure down to the resolution capability of the instrument
of $0.1''$, while isolated point-like brightenings exist as well. The structure made 
up of these objects evolves on a shorter than granular time-scale, giving the 
impression of a separate (magnetic) fluid that resists mixing with the granular 
material. Figure~\ref{steiner_fig01} shows an example G-band filtergram from
the former paper taken in a remnant active region plage near disk center.
In this region, intergranular lanes are often completely filled with 
magnetic field like in the case marked by the white lines in 
Fig.~\ref{steiner_fig01}. 
There, and in other similar cases, the magnetic field concentration is framed 
by a striation of bright material, while the central part is dark. 
Figure~\ref{steiner_fig01} shows examples of ribbon bands and
also an isolated bright point in the lower right corner.

The graphic to the right hand side of Fig.~\ref{steiner_fig01} displays the 
emergent G-band intensity (solid curve) from the cross section marked by 
the white horizontal lines in the image to the left. Also shown are the 
corresponding magnetographic signal 
(dashed curve), the blue continuum intensity (dotted), and the Ca H-line
intensity (dash-dotted). Note that the magnetic signal is confined to the
gap between the two horizontal white lines. The intensities show
a two-humped profile.

\begin{figure}
\centering
\includegraphics[width=1.0\linewidth]{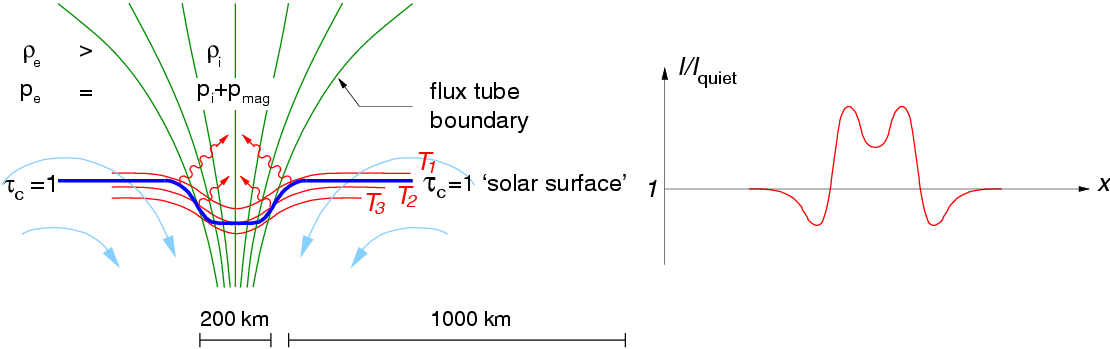}
\caption{Sketch of a magnetic flux sheet (left) with corresponding
intensity contrast (right), distilled from self-consistent numerical
MHD simulations. Note that the isothermal surfaces are not 
exactly parallel to the surface of optical depth unity, which gives rise
to the particular M-shape of the contrast profile.
\label{steiner_fig02}
}
\end{figure}

This situation reminds of the flux-sheet model and the 
``bright wall effect''. A first quasi-stationary, self-consistent 
simulation of a small-scale flux sheet was carried out by 
\citet{deinzer+al84a,deinzer+al84b}, popularly known as the ``KGB-models''.
The basic properties of this model are sketched in
Fig.~\ref{steiner_fig02}. Accordingly, a small-scale flux concentration, either 
of tube or sheet-like shape, is in mechanical equilibrium with the ambient
atmosphere, viz., the gas plus magnetic pressure of the atmosphere within the 
tube/sheet balances the gas pressure in the ambient (field-free) medium at equal 
geometrical height. This situation calls for a reduced
density in the flux concentration with respect to 
the environment, at least in the photospheric part, where the radiative heat 
exchange quickly drives the configuration towards radiative equilibrium, hence 
to a similar temperature at constant geometrical height. This density reduction 
renders the flux tube/sheet atmosphere  more 
transparent, which causes a depression of the surface of constant  optical depth,
as indicated by the surface of $\tau_c = 1$ in Fig.~\ref{steiner_fig02}. In a plage or
network region, this effect increases the ``roughness'' of the solar surface,
hence the effective surface from which radiation can escape, which increases
the net radiative loss from these areas. 

The graphics to the right hand side of Fig.~\ref{steiner_fig02}  shows a sketch of
the relative intensity emerging from this model, viz., the intensity of light 
propagating in the vertical direction as a function of distance from the  
flux sheet's plane of symmetry.
It corresponds to the plot on the right hand side of Fig.~\ref{steiner_fig01}.
The similarity between this model  and  the observation is striking.  
Turning to a narrower flux sheet/tube would result in the merging of the two contrast 
peaks to a single central peak in both, model and observation, i.e., to a ribbon
band or bright point, respectively.
Yet, the striation of the depression wall that can be seen in the observation 
is of course not reproduced by the model, which is strictly two-dimensional with 
translational invariance in lane direction. We will see in Sect.~\ref{sect_fac}
that three-dimensional magnetoconvection
simulations show rudimentary striation. The physical origin of the striation is still 
unknown.

Accordingly, the basic properties of ribbon-like magnetic flux concentrations 
can be understood in terms of a magnetic flux sheet embedded in and in force 
balance with a more or less field-free ambient medium. This can also be said
(replacing the word sheet by tube) of the rosette structure visible in other
still images of \citet{berger+al04} who call it ``flower-like''. Flowers can transmute 
to pores and vice versa. The striation of their bright collar is similar
to that seen in ribbon structures. Discarding the striation, the basic 
properties of flowers can well be interpreted in terms of a tube shaped 
flux concentration like the one sketched in Fig.~\ref{steiner_fig02}.

A 2~h sequence of images with a quality comparable to  Figs.~\ref{steiner_fig01} 
\citep{vandervoort05} reveals that the shape of the
ribbon-like flux concentrations and the striation of ribbons and flowers change 
on a very short time-scale, of the order of the Alfv\'en crossing travel time.
This suggests that these morphological changes and the striation itself are
related to the flute instability, which small-scale flux concentrations are liable 
to. For an untwisted axisymmetric flux tube, the radial component of the
magnetic field at the flux-tube surface must decrease with height, 
$\left. {\rm d} B_r /{\rm d} z\right|_S < 0$, 
in order that the flux tube
is stable against the flute (interchange) instability \citep{meyer+schmidt77}.
While sunspots and pores with a magnetic flux in excess of 
$\Phi \approx 10^{19}$\,Mx do meet this condition, small-scale flux concentrations 
do not fulfill it \citep{schuessler84,steiner_phd,buente+al93}. 
\citet{buente93} shows that small-scale flux sheets too are flute unstable,
and he concludes that filament formation due to the flute instability close
to the surface of optical depth unity would ensue. As the flux sheet is bound 
to fall apart because of the flute instability, its debris are 
again reassembled by the continuous
advection back to the intergranular lane so that a competition between the 
two effects is expected to take place, which might be at the origin of the 
corrugation of the field concentrations and of the striation of the 
tube/sheet interface with the ambient medium.

Although the fine structure of small-scale magnetic flux concentrations 
changes on a very short time scale, single flux elements seem to persist
over the full duration of the time sequence of 2 h. They may dissolve
or disappear for a short period of time, but it seems that the same magnetic
flux continually reassembles to make them reappear nearby. Latest G-band 
time sequences obtained with the Solar Optical Telescope (SOT) on board of the
Japanese space satellite HINODE ({\tt http://solarb.msfc.nasa.gov/movies.html})
seem to confirm these findings even for G-band bright points of low intensity.
This suggest a deep anchoring of at least some of the flux elements
although numerical simulation seem not to confirm this conjecture.

As indicated in the sketch of Fig.~\ref{steiner_fig02}, the magnetic 
flux concentration is framed by a downflow of material, fed by a horizontal 
flow that impinges on the flux concentration. Already the flux-sheet model 
of \citet{deinzer+al84b} showed a persistent flow of this kind. According to
these authors it is due to radiative cooling from the depression 
walls of the magnetic flux concentrations (the ``hot wall effect'') that  causes
a horizontal pressure gradient, which drives the flow. The non-stationary
flux-sheet simulations of \citet{steiner+al1998} and \citet{leka+steiner2001} 
showed a similar persistent downflow, which, with increasing depth, becomes faster 
and narrower, turning into  veritable {\it downflow jets} beneath the visible
surface. While downflows in the periphery of pores have been observed
earlier \citep{leka+steiner2001,sanka+rimmele03,tritschler+al03}
and also horizontal motions towards a pore by \citet{dorotovic+al02},
only very recently such an accelerating downflow has been observationally 
detected in the immediate vicinity of ribbon bands by \citet{langangen+al07}.


\section{3-D simulations of small-scale magnetic flux concentrations}
\label{sect_sim}

New results from realistic simulations on the formation, dynamics and structure 
of small-scale magnetic flux concentrations have recently been published 
in a series of papers by Sch\"ussler and collaborators.
\citet{voegler+al05} simulate magnetoconvection in a box encompassing
an area on the solar surface of $6\times 6$~Mm$^2$ with a height extension
of 1400~km, reaching from the temperature minimum to 800~km below the
surface of optical depth unity. Although this is only 
0.4\% of the convection zone depth, the box still includes the 
entire transition from almost completely convective to mainly radiative energy
transfer and the transition from the regime where the flux concentration is dominated 
by the convective plasma flow to layers where the magnetic energy density 
of the flux concentrations by far surpasses the thermal energy density.
The bottom boundary in this and similar simulations is open in the sense
that plasma can freely flow in and out of the computational domain, subject
to the condition of mass conservation. Inflowing material has a given specific
entropy that determines the effective temperature of the radiation leaving
the domain at the top, while the outflowing material carries the entropy it
instantly has.

\begin{figure}
\centering
\includegraphics[width=0.40\linewidth]{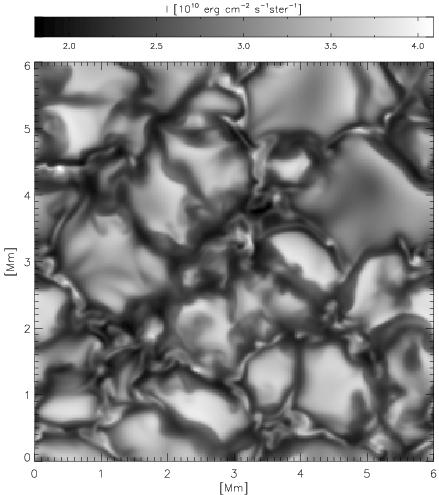}\hspace{0.05\linewidth}
\includegraphics[width=0.40\linewidth]{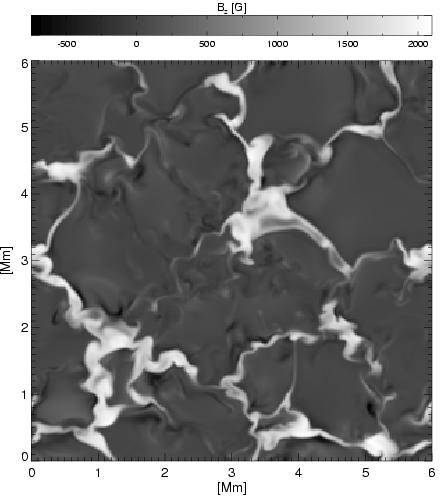}
\caption{Simulation snapshot. Left: Frequency integrated intensity. Right:
Vertical magnetic flux density at constant average geometrical height of 
optical depth unity. The mean flux density is 200~G. From \citet{voegler+al05}.
\label{steiner_fig03}
}
\end{figure}

Figure~\ref{steiner_fig03} shows a snapshot from this simulation: To the left the
emergent mean intensity, to the right the vertical magnetic field strength
at a constant height, viz., at the horizontally averaged geometrical height of 
optical depth unity. (I would like to caution that this magnetic map is not
what would be seen with a magnetograph, irrespective of its spatial resolution.)
The strong magnetic field in intergranular lanes is manifest
in a corresponding signal in the emergent intensity very much like
the snapshot discussed in Sect.~\ref{sect_obs}. Also the intensity signal
shows the same corrugated and knotted ribbon structure that is observed, and 
sometimes there appear also broader ribbon structures with a dark central core,
like the one marked in Fig.~\ref{steiner_fig01}.
In the latter case however, the characteristic striation is absent, possibly because 
the flute instability is suppressed on very small spatial scales due to lack 
of sufficient resolution of the simulation. In the central part of the
snapshot, a micro pore or magnetic knot has formed.

A comparison of the average gas plus magnetic pressure as a function of height
at locations of magnetic flux concentrations with the run of the average gas 
pressure in weak-field regions reveals that the two are almost identical, proving
that even in this dynamic regime, the thin flux tube approximation is very well 
satisfied \citep{voegler+al05}. This result  confirms that
the model discussed in Sect.~\ref{sect_obs} and sketched in Fig.~\ref{steiner_fig02} 
is indeed an acceptable first approximation to the real situation.

Simulations are not just carried out for the sake of reproducing observed
quantities. Once good agreement with all kind of observations exists, simulations
allow with some confidence to inform about regions not directly accessible
to observations, for example about the magnetic structure in subsurface
layers. In this respect the simulations of \citet{voegler+al05} show that often 
flux concentrations that have formed at the surface disperse again in shallow depths.
This behaviour was also found by \citet{schaffenberger+al05} in
their simulation with an entirely different code and further by
\citet{stein+nordlund06}.   A vertical section through a 
three-dimensional simulation domain of \citet{schaffenberger+al05}, where two such
shallow flux concentrations have formed, is shown in Fig.~\ref{steiner_fig09}.
The superficial nature of magnetic flux concentrations in the simulations,
however, is difficult to reconcile with the observation that
many flux elements seem to persist over a long time period.


\section{The physics of faculae}
\label{sect_fac}

With growing distance from disk center, small-scale 
magnetic flux concentrations grow in contrast against the quiet 
Sun background and become apparent as solar faculae close to the limb. 
Ensembles of faculae form plage and network faculae that 
are as conspicuous features of the white light solar disk, 
as are sunspots. It is therefore not surprising that they play a key 
role in the solar irradiance variation over a solar cycle and on 
shorter time scales \citep{fligge+al00,wenzler+al05,foukal+al06}.
Measurements of the center to limb variation of the continuum contrast 
of faculae are diverse,
however, as the contrast is not only a function of  the heliocentric
angle, $\mu = \cos\theta$, but also of facular size, magnetic field
strength, spatial resolution, etc., and as measurements are prone to selection 
effects. While many earlier measurements report a contrast maximum around
$\mu \approx 0.2\ldots 0.4$ with a decline towards the limb, latest measurements
\citep{sutterlin+al99,ahern+chapman00,adjabshirizadeh+koutchmy02,
ortiz+al02,centrone+ermolli03,vogler+al05} point rather to a monotonically 
increasing or at most mildly decreasing contrast out to the limb.

The standard facula model \citep{spruit76}, again consists of a magnetic flux 
concentration embedded in and in mechanical equilibrium with a weak-field 
or field-free environment as is sketched in Fig,~\ref{steiner_fig02}. When
approaching the limb, the limb side of the bright depression wall becomes ever 
more visible and ever more perpendicular to the line of sight, which increases its
brightness compared to the limb darkened environment. At the extreme limb, 
obscuration by the centerward rim of the depression starts to take place, 
which decreases the size and possibly the contrast of the visible limb-side wall.

\begin{figure}
\centering
\begin{minipage}[b]{0.40\textwidth}
\includegraphics[width=1.0\linewidth]{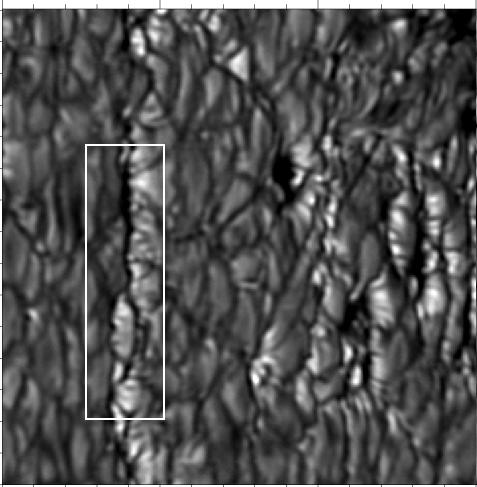}
\end{minipage}\hspace{0.05\linewidth}
\begin{minipage}[b]{0.45\textwidth}
\includegraphics[width=1.0\textwidth]{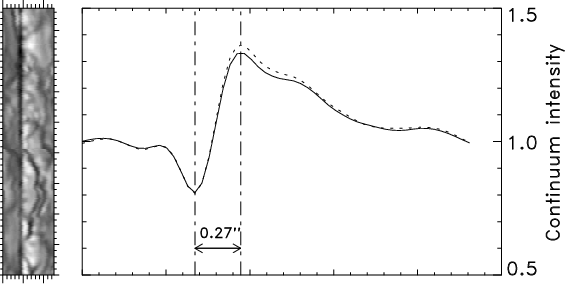}
\vspace*{0.05\linewidth}
\end{minipage}
\caption{Left: Network faculae at a heliocentric angle of $\mu = 0.48$ in the continuum 
at 587.5~nm. The solar limb is to the right.
Right: Faculae within the white box of the image to the left aligned according to the 
position of the dark lane, together with the mean spatial scan through the aligned faculae.
From \citet{hirzberger+wiehr05}.
\label{steiner_fig04}
}
\end{figure}

Recently, \citet{lites+al04} and \citet{hirzberger+wiehr05} have obtained excellent
images of faculae with the 1~m Swedish Solar Telescope. Figure~\ref{steiner_fig04}
(from  \citealt{hirzberger+wiehr05}) shows on the left hand side network faculae 
at a heliocentric angle of $\mu = 0.48$ in the continuum at 587.5~nm. 
The solar limb is located 
towards the right hand side. It is clearly visible from this image that faculae are 
in reality partially brightened granules with an exceptionally dark and wide
intergranular lane (``dark facular lane'') on the disk-center 
side of the contrast enhancement, which is also the location of the magnetic flux
concentration. The right half of Fig.~\ref{steiner_fig04} shows the string of faculae
within the white box of the image to the left, aligned according to the position
of the dark lane. Also shown is the mean contrast profile, averaged over the alignment. 
Similar contrast profiles of single faculae are shown by \citet{lites+al04}. Such 
contrast profiles pose now a new constraint that any model of faculae must satisfy.

Magnetoconvective simulations as the one discussed in Sect.~\ref{sect_sim}
indeed show facular-like contrast enhancements when computing the
emergent intensity along lines of sight that are inclined to the vertical direction
for mimicking limb observation. Such tilted three-dimensional simulation 
boxes are shown in the
papers by \citet{keller+al04}, \citet{carlsson+al04}, and \citet{depontieu+al06}. 
\citet{keller+al04} also show the contrast profile of two isolated ``faculae'', 
which however have a more symmetric shape, rather than the observed characteristic 
steep increase on the disk-center side with the gentle decrease  towards the limb. 
Also they obtain a maximum contrast of 2, far exceeding the observed value of 
about 1.3. It is not clear what the reason for this discrepancy might be. 
Interestingly, the old ``KGB-model'' \citep{deinzer+al84b,knoe+msch88} as well as 
the two-dimensional, non-stationary simulation of \citet{steiner05}
do nicely reproduce the asymmetric shape and the dark lane.

Another conspicuous property of faculae that high-resolution images reveal is that
they are not uniformly bright but show a striation not unlike to and possibly
in connection with the one seen in G-band ribbons at disk center. While this 
feature cannot be reproduced in a two-dimensional model, it must be part of a
satisfactory three-dimensional model. But so far 3-D simulations show only
a rudimentary striation. This finding, rather surprisingly, indicates that the 
effective spatial resolution of present-day three-dimensional simulations is 
inferior to the spatial resolution of best current observations.

\begin{figure}
\centering
\includegraphics[width=1.0\linewidth]{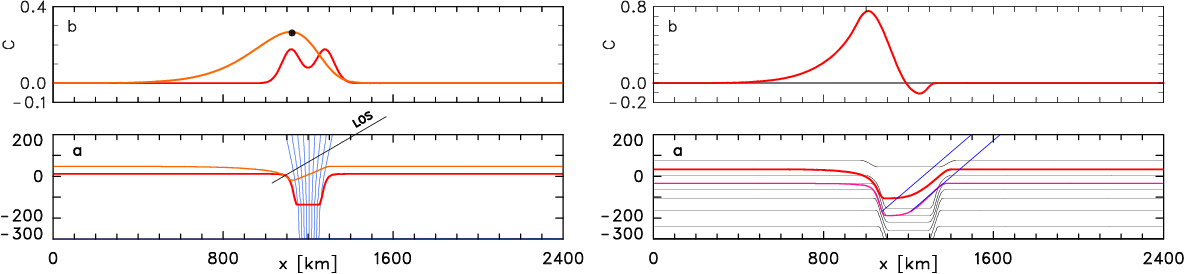}
\caption{
Left: a) Magnetic flux concentration (blue, vertically oriented lines of force, 
see internet version for colours) 
with surfaces of optical depth $\tau = 1$ for vertical lines of sight 
(thick/red curve) and lines of sight inclined by $\theta = 60^{\circ}$ to the
vertical direction (thin/red curve). b) Corresponding contrast curves. 
All values of the light red curve left of the black dot originate from lines 
of sight left of the one indicated in panel a).
Right:  a) Surfaces of optical depth $\tau = 1$ and $5$ (thick/red) for 
lines of sight inclined by $50^{\circ}$ to the vertical, together with isotherms. 
b) Contrast profile. The region of negative contrast is bounded by the 
two oblique lines of sight indicated in panel a).
\label{steiner_fig05}
}
\end{figure}

\begin{figure}
\centering
\includegraphics[width=0.4\linewidth]{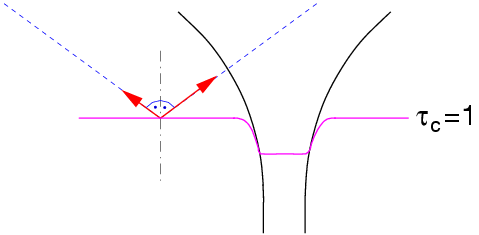}
\caption{Photons preferentially escape along the line of sight to the right that
traverses the magnetic flux sheet/tube in comparison to the line of sight to the
left under equal zenith angle, because of the rarified (less opaque) atmosphere
in the flux sheet/tube.
\label{steiner_fig06}
}
\end{figure}

In an attempt to better understand the basic properties of faculae,
\cite{steiner05} considers the ideal model of a magnetohydrostatic flux sheet 
embedded in a plane parallel standard solar atmosphere. For the construction
of this model the flux-sheet atmosphere is first taken to be identical to the 
atmosphere of the ambient medium but shifted in the 
downward direction by the amount of the ``Wilson depression'' (the depression of the 
surface of continuum optical depth unity at the location of the flux concentration).
The shifting results in a flux-tube atmosphere that is less dense and cooler than
the ambient medium at a fixed geometrical height.
In the photospheric part of the flux concentration, thermal equilibrium with the 
ambient medium is then enforced. Denoting with index $i$ the flux-sheet
atmosphere and with $e$ the ambient atmosphere and with $W$ the depth of the
``Wilson depression'', we therefore have
\begin {equation}
T_i(z) = \!\!\left\{ \!\!\! \begin{array}{l}
T_e(z+W)\quad \mbox{for}\quad \tau_c \gg 1\\
T_e(z)\quad \mbox{for}\quad \tau_c \ll 1\;,
\end{array}\right.
\end {equation}
where $\tau_c$ is the optical depth in the visible continuum and
$\rho_i(z) < \rho_e(z)\; \forall z$. The lower left panel
of Fig.~\ref{steiner_fig05} shows this configuration together with two surfaces of
optical depth unity, one for vertical lines of sight (disk center), the other for 
lines of sight running from the top right to the bottom left under an angle of 
$\theta = 60^{\circ}$ to the vertical, like the one indicated in the figure. 
The upper left panel shows the corresponding continuum enhancement for 
disk center (double humped profile) and 
$\theta = 60^{\circ}$. Of the curve belonging to $\theta = 60^{\circ}$, all values
left of the black dot belong to lines of sight left of the one indicated in the lower 
panel. This means that the contrast enhancement extends far beyond the
depression proper in the limbward direction, exactly as is observed. The
reason for this behaviour is explained with the help of Fig.~\ref{steiner_fig06} 
as follows.

A material parcel located in the solar atmosphere and lateral to the flux sheet ``sees'' 
a more  transparent atmosphere in the direction toward the flux sheet compared 
to a direction under equal zenith angle but pointing away from it because of
the rarefied flux-sheet atmosphere. Correspondingly, from a wide area 
surrounding the magnetic flux sheet or flux tube, radiation escapes more 
easily in the direction  towards the flux sheet so that a single flux sheet/tube impacts 
the radiative escape in a cross-sectional area that is much wider than the 
magnetic field  concentration proper.

The right hand side of Fig.~\ref{steiner_fig05} shows a similar situation as to
the left but for a flux sheet that is twice as wide. The continuum contrast
for lines of sight inclined by $\theta = 50^{\circ}$ to the vertical is shown in the
top panel. It can be seen that a dark lane of negative contrast 
 occurs on the disk-center side of the facula.
It arises from the low temperature gradient of the flux-sheet atmosphere in the 
height range of $\tau_c = 1$ and its downshift relative to the external atmosphere in 
combination with the inclined lines of sight. One could say that the dark lane 
in this case is an expression of the cool ``bottom'' of the magnetic flux sheet. 

It is remarkable that this basic, energetically not self-consistent
model is capable of producing both, the facular dark lane and the asymmetrically
shaped contrast curve of the facula, with realistic contrast values. The
results of this basic hydrostatic model carry over to a fully self-consistent
model of a magnetic flux sheet in dynamic interaction with non-stationary 
convective motion \citep{steiner05}. In this case the facular lane becomes 
broader and darker.

It follows from these insights that a facula is not to be identified with 
bright plasma that sticks,  as the name may insinuate, like a torch out of 
the solar surface and as the ``hillock model'' of \citet{schatten+al86} 
suggests. Rather is it the manifestation of photospheric granulation, seen 
across a magnetic flux concentration --- granulation that appears brighter 
than normal in the form of so called ``facular granules''.
Interestingly, already \citet{chevalier1912} wondered: ``La granulation que l'on 
voit autour des taches plus \'eclatante que sur les autres parties est-elle la 
granulation des facules ou celle de la photosph\`ere vue \`a travers les facules ?''
and \citet{bruggencate40} noted that ``Sie [Photosph\"arengranulen und Fackelgranulen]
unterscheiden sich nicht durch ihre mittlere Gr\"osse, wohl aber durch den Kontrast 
gegen\"uber der Umgebung.''

If this is true, one expects facular granules to show the same dynamic phenomena like
regular granulation. Indeed, this is confirmed in a comparison of observations with
three-dimensional simulations by \citet{depontieu+al06}. 
They observe that often a dark band gradually moves from the limb side of a facula
toward the disk center and seemingly sweeps over and "erases" the facula temporarily.
The same phenomenon they also observed in a time sequence of a three-dimensional 
simulation, which enabled them to identify the physics behind this phenomenon.

Examination of the simulation sequence reveals that dark bands are a consequence of
the evolution of granules. Often granules show a dark lane that usually introduces
fragmentation of the granule. The smaller fragment often dissolves (collapses) in 
which case the dark lane disappears with the collapsing small fragment in the 
intergranular space. Exactly this process can lead to the dark band phenomenon,
when a granular dark lane is swept towards the facular magnetic flux concentration.
Since the facular brightening is seen in the disk-center
facing side of granules, only granular lanes that are advected in the direction of
disk center lead to facular dimming. This example nicely demonstrates how regular 
granular dynamics when seen across the facular magnetic field can lead to 
genuine facular phenomena.
 
Despite the major progress that we have achieved in understanding the physics of 
faculae over the past few years, open questions remain. These concern
\begin{list}{-}{\setlength{\topsep}{3pt}\setlength{\parsep}{0.0mm}
                      \setlength{\itemsep}{3pt}\setlength{\leftmargin}{10pt}}
\item[-] a comprehensive model of the center to limb 
         variation of the brightness of faculae including dependence on size, magnetic flux,
         flux density, color, etc.,
\item[-] a quantitative agreement between simulation and observation with respect to
         measurements in the infrared and with respect to the observed
         geometrical displacement between line core and continuum filtergrams of 
         faculae,
\item[-] the physical origin of the striation, 
\item[-] a quantitative evaluation of the heat leakage caused by faculae, or
\item[-] the role of faculae in guiding magnetoacoustic waves into the chromosphere.
\end{list}


\section{3-D MHD simulation from the convection zone to the chromosphere}
\label{sect_3dmhd}

For investigating the connection between photospheric small scale magnetic fields
and the chromosphere, \citet{schaffenberger+al05} have extended the three-dimensional
radiation hydrodynamics code
\cobold\ \footnote{www.astro.uu.se/\~{}bf/co5bold\_main.html}
to magnetohydrodynamics for studying
magnetoconvective processes in a three-dimensional environment
that encompasses the integral layers from the top of the convection
zone to the mid chromosphere. The code is based on a finite volume scheme,
where fluxes are computed with an approximate Riemann-solver 
\citep{leveque+al99,toro99} for automatic shock capturing. For the advection 
of the magnetic field components, a constrained transport scheme is used.

The three-dimensional computational domain extends from 1400~km below
the surface of optical depth unity to 1400~km above it and it has a horizontal
dimension of $4800\times 4800$~km. The simulation starts with a homogeneous,
vertical, unipolar magnetic field of a flux density of 10~G superposed on a 
previously computed, relaxed model of thermal convection. This low flux density 
is representative for magnetoconvection in a very quiet network-cell interior.
The magnetic field is constrained to have vanishing
horizontal components at the top and bottom boundary, but lines of force
can freely move in the horizontal direction, allowing for flux concentrations
to extend right to the boundaries. Because of the top boundary being
located at mid-chromospheric heights, the magnetic field is allowed to freely 
expand with height through the photospheric layers into the more or less
homogeneous chromospheric field.

\begin{figure}
\centering
\begin{minipage}[b]{0.70\textwidth}
  \includegraphics[width=1.0\textwidth]{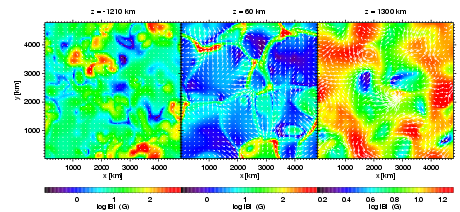}
\end{minipage}\hspace*{0.01\textwidth}
\begin{minipage}[b]{0.275\textwidth}
  \includegraphics[width=1.0\linewidth]{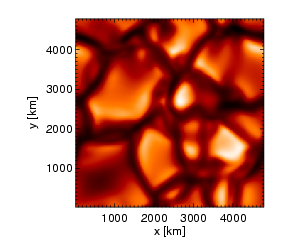}
  \vspace*{0.07\textwidth}
\end{minipage}
\caption{Horizontal sections through the three-dimensional computational domain. 
Color coding displays $\log |B|$ with individual scaling for each panel
(see internet version for colours).  Left: 
Bottom layer at a depth of 1210~km. Middle: Layer 60~km above optical depth 
$\tau_c = 1$. Right: Top, chromospheric layer in a height of 1300~km. White 
arrows indicate the horizontal velocity on a common scaling. 
Longest arrows in the panels from left to right correspond to
4.5, 8.8, and 25.2 km/s, respectively. 
Rightmost: Emergent visible continuum intensity.
From \citet{schaffenberger+al05}.
\label{steiner_fig08}
}
\end{figure}

\begin{figure}
\centering
 \includegraphics[width=1.0\linewidth]{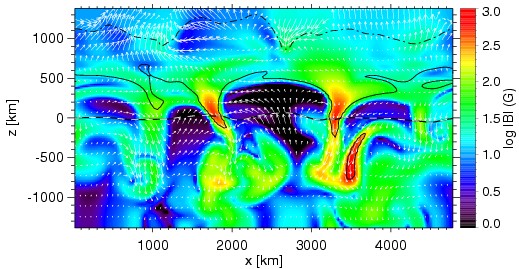}
\caption{Snapshot of a vertical section through the three-dimensional 
computational domain, showing $\log |B|$ (color coded) and velocity 
vectors projected on the vertical plane (white arrows). The b/w dashed 
curve shows optical depth unity and the dot-dashed and 
solid black contours $\beta = 1$ and 100, respectively. 
See internet version for colours. From \citet{schaffenberger+al05}.
\label{steiner_fig09}
}
\end{figure}

Figure~\ref{steiner_fig08} shows the logarithmic
absolute magnetic flux density in three horizontal sections through the 
computational domain at a given time instant, together with the emergent
Rosseland mean intensity.
The magnetic field in the chromospheric part is marked by strong dynamics 
with a continuous rearrangement of magnetic flux on a time scale of
less than 1~min, much shorter  than in the photosphere or in the
convection-zone layers. There, the field has a strength between 2 and 40 G
in the snapshot of Fig.~\ref{steiner_fig08}, which is typical for the whole 
time series. Different from the surface magnetic field, it is more 
homogeneous and practically fills the entire space so that the magnetic 
filling factor in the top layer is close to unity. There seems to be no 
spatial correlation between chromospheric flux concentrations and 
the small-scale field concentrations in the photosphere. 

\begin{figure}
\centering
 \includegraphics[width=0.9\linewidth]{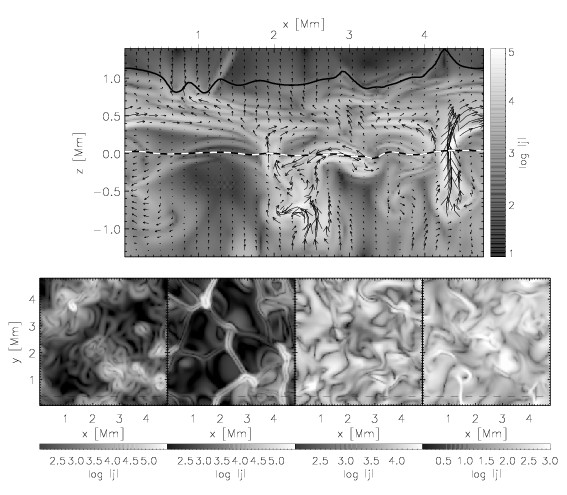}
\caption{Logarithmic current density, $\log |j|$, in a vertical cross 
   section (top panel) and in four horizontal cross sections in a depth of 
   1180\,km below, and at heights of 90\,km, 610\,km, and 1310\,km above the 
   mean surface of optical depth unity from left to right, respectively.
   The arrows in the top panel indicate the magnetic field strength and
   direction. The dashed line indicates the position of the vertical section.
   [$j$] = $3\,\times\,10^5$ A/m$^{2}$. From \citet{wedemeyer+al07}.
   \label{steiner_fig10}
}
\end{figure}

Comparing the flux density of the panel corresponding to $z = 60$~km with 
the emergent intensity, one readily sees that the magnetic
field is concentrated in intergranular lanes and at lane vertices. However,
the field concentrations do not manifest a corresponding intensity signal
like in Fig.~\ref{steiner_fig03}. This is because the magnetic flux is too weak 
to form a significant ``Wilson depression'' (as can be seen from 
Fig.~\ref{steiner_fig09}) so that no radiative channeling effect takes place.

Figure~\ref{steiner_fig09} shows the logarithm of the absolute field strength 
through a vertical section of the computational domain. 
Overplotted are white arrows indicating the velocity field. The b/w dashed
curve corresponds to the optical depth unity for vertical lines of
sight. Contours of the ratio of thermal to magnetic pressure, 
$\beta$, for $\beta = 1$ (dot-dashed) and $\beta = 100$ (solid) are also 
shown. Magnetoacoustic waves that form transient filaments of stronger 
than average magnetic field are a ubiquitous phenomenon in the 
chromosphere and are also present in the snapshot of 
Fig.~\ref{steiner_fig09}, e.g., 
along the contour of $\beta = 1$ near $x = 1200$~km and $x = 2500$~km.
They form in the compression zone downstream and along propagating 
shock fronts. These magnetic filaments that have a field strength rarely 
exceeding 40~G, rapidly move with the shock fronts and quickly form 
and dissolve with them.  

The surface of $\beta = 1$ separates the region of highly dynamic magnetic fields 
around and above it from the more slowly evolving field of high beta plasma below
it. This surface is located at approximately 1000~km but it is corrugated 
and its local height strongly varies in time.

A very common phenomenon in this simulation is the formation of a
`magnetic canopy field' that extends in a more or less horizontal
direction over expanding granules and between photospheric flux
concentrations. The formation of such canopy fields proceeds by the
action of the expanding flow above granule centres. This flow
transports `shells' of horizontal magnetic field to the upper
photosphere and lower chromosphere, where shells of different field
directions may be pushed close together, leading to a complicated
network of current sheets in a height range from approximately 400 to
900\,km.

This network can be seen in Fig.~\ref{steiner_fig10} (top), which shows, for
a typical snapshot of the simulation, the logarithmic current density,
$\log |j|$, 
together with arrows indicating the magnetic field strength and direction. 
Figure~\ref{steiner_fig10} (bottom) shows from left to right 
$\log |j|$ 
in four horizontal cross sections in a depth of
1180\,km below, and at heights of 90\,km, 610\,km, and 1310\,km above
the mean surface of optical depth unity. Higher up in the chromosphere
(rightmost panel), thin current sheets form along shock fronts, e.g.,
in the lower left corner near \mbox{$x = 1.4$\,Mm}.

Using molecular values for the electrical conductivity, 
\citet{wedemeyer+al07} derive an energy flux of 5 to 50\,W\,m$^{-2}$
into the chromosphere caused by ohmic dissipation of these 
current sheets. This value is about two orders of magnitude short 
of being relevant for chromospheric heating. On the other hand,
the employed molecular values for the conductivity might be orders
of magnitude too high for to be compatible with the effective 
electrical conductivity of the numerical scheme determined 
by the inherent artificial diffusion. Therefore, the ohmic heat flux
might be conceivably two orders of magnitude larger than suggested by
this rough estimate, so that magnetic heating by ohmic dissipation 
must be seriously taken into account. More advanced simulations, taking 
explicit ohmic diffusion into account will clarify this issue.


\section{Future directions}

Continuously increasing power of computational facilities together
with steadily improving computational methods, expand 
the opportunity for numerical simulations. On the one hand, 
more detailed physics can be included in the simulation, on the other 
hand either the computational domain or the
spatial and temporal resolution can be increased. 
In most simulations, especially when the computational domain encompasses
only a small piece of a star, 
boundary conditions play an important role. They convey information on the 
outside world to the physical domain of the simulation. But this 
outside world is often poorly known. In order to acquire experience 
and intuition with respect to the influence of different types of boundary 
condition on the solution, one can implement and run various realizations 
of boundary conditions, which however also requires additional resources 
in computer power and time. Also the initial condition may critically 
determine the solution, for example the net flux and flux density of an 
initial, homogeneous vertical magnetic field. Boundary conditions, therefore, 
remain a hot topic also in future.

Most excitement in carrying out numerical simulations comes 
from the prospect of performing experiments with the object under
investigation: experiments in the numerical laboratory.
Not only that an astrophysical object can be reconstructed and simulated
in the virtual world of the numerical laboratory. Once in the computer, the
computational astrophysicist has the prospect of carrying out experiments 
with it as if the celestial body was taken to the laboratory.

The following few examples shall illustrate some aspects of this.

\subsection{More detailed physics}

In the solar chromosphere the assumption of LTE (local thermodynamic equilibrium) 
is not valid. Even the assumption of statistical equilibrium in the rate equations 
is not valid because the relaxation time-scale for the ionization of hydrogen 
approaches and surpasses the hydrodynamical time scale in the chromosphere 
\citep{kneer80}. Yet, in order to take time dependent hydrogen 
ionization in a three-dimensional simulation into account, simplifications are 
needed. \citet{leenaarts+wedemeyer06} employ the 
method of fixed radiative rates for a hydrogen model atom with six energy levels 
in the three-dimensional radiation (magneto-)hydrodynamics code \cobold. Thus,
additional to the hydrodynamic equations, they solve the time-dependent rate equations
\begin{equation}
\frac{\partial n_i}{\partial t} + \nabla\cdot(n_i \mbox{\boldmath v}) = 
\sum_{j\ne i}^{n_l} n_j P_{ji} - n_i \sum_{j\ne i}^{n_l} P_{ij}\,,
\end{equation}
with $P_{ij}$ being the sum of collisional and  radiative rate coefficients,
$P_{ij} = C_{ij} + R_{ij}$. The rate coefficients are now local quantities
given a fixed radiation field for each transition, which is obtained from 
one-dimensional test calculations.

Simulations with this approach show that above the height of the classical 
temperature minimum, the non-equilibrium ionization degree is fairly constant 
over time and space at a value set by hot propagating shock 
waves. This is in sharp contrast to results with LTE, where
the ionization degree varies by more than 20 orders of magnitude between 
hot gas immediately behind the shock front and cool regions further away. 
The addition of a hydrogen model atom provides realistic values for hydrogen 
ionization degree and electron density, needed for detailed radiative transfer 
diagnostics.

\subsection{Large box simulations}

\citet{benson+al07} have carried out first simulations with a large
simulation box so as to accommodate a supergranulation cell. They
started a simulation that encompasses a volume of $48 \times 48\times 20$~Mm$^3$ 
using $500^3$ grid cells. With this simulation they hope to find out more
about the origin of supergranulation and to carry out helioseismic
experiments \citep{zhao+al07}.

\citet{hansteen04} has carried out MHD simulations comprising a vast height 
range from the top layers of the convection zone into the transition region and
the corona. With these simulations they seek to investigate various
chromospheric features such as dynamic fibrils  \citep{hansteen+al06},
mottles, and spicules, which are some of the most important, but also most 
poorly understood, phenomena of the Sun's magnetized outer atmosphere.
But also the transition zone and coronal heating mechanisms are in
the focus of these kinds of ``holistic'' simulations.

\begin{figure}
\centering
\includegraphics[width=0.48\linewidth]{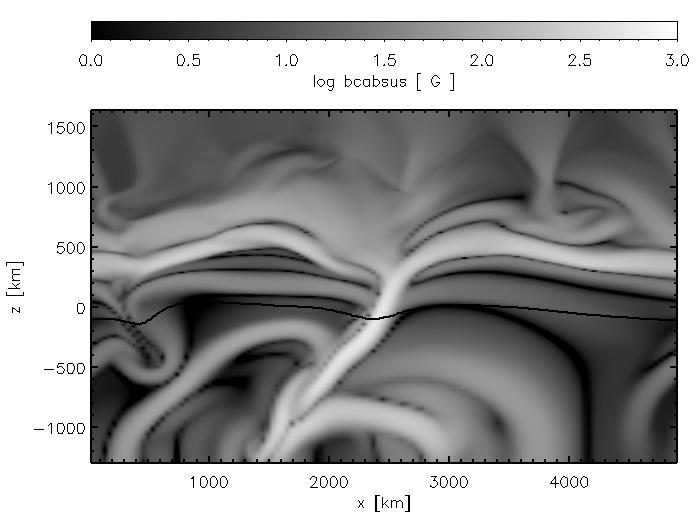}\hfill
\includegraphics[width=0.48\linewidth]{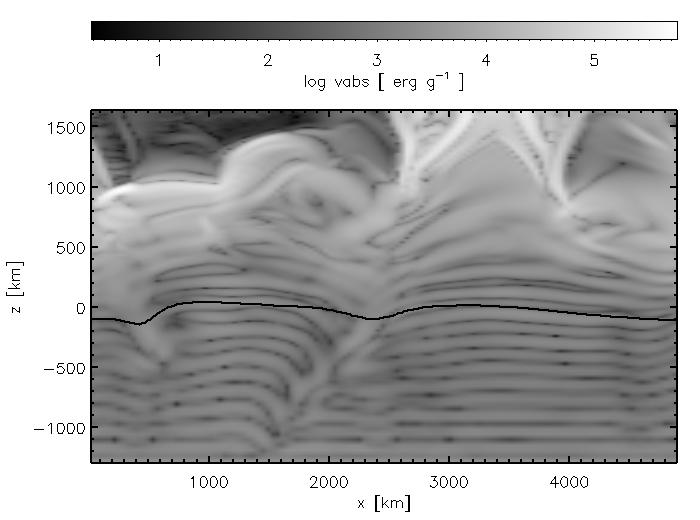}
\caption{
Left: Still image of the logarithmic magnetic flux density from a time series for
the instant $t=1368$ s after starting with an initial homogeneous vertical field 
of 10~G flux density. A strong magnetic flux sheet has formed extending from
$(x,z)\approx (2000,-500)$ to  $(x,z)\approx (2500,0)$.
Right: A plane parallel wave with frequency 100 mHz travels through convecting 
plasma into the magnetically structured photosphere and further into the low 
$\beta$ (magnetically dominated) chromosphere. The panel shows the difference 
in absolute velocity between the perturbed and the unperturbed solution 212~s
after the start of the perturbation. The wave becomes strongly refracted
in the low $\beta$ region and at the location of the flux sheet.
\label{steiner_fig11}
}
\end{figure}

\subsection{Improvements in boundary conditions}

Many conventional magnetohydrodynamic simulations of the small-scale solar 
magnetic field assume that the horizontal component of the magnetic field 
vanishes at the top and bottom of the computational domain 
\citep[e.g.][]{weiss+al96,cattaneo+al03,voegler+al05,schaffenberger+al05},
which is a rather stark constraint, especially with respect to 
magnetoacoustic wave propagation and Poynting flux.
Recently, \citet{stein+al06} have introduced an alternative
condition with the possibility of advecting magnetic field across the bottom 
boundary. Thus, upflows into the computational domain carry horizontal
magnetic field of a prescribed flux density with them, while outflowing
plasma carries whatever magnetic field it instantly has.
With this condition an equilibrium in which equal amounts of
magnetic flux are transported in and out of the computational domain is
approached after some time. It should more faithfully
model the plasma flow across the lower boundary and
it allows for the effect of magnetic pumping \citep{tobias+al98}.

\subsection{Helioseismic experiment with a magnetically structured atmosphere}

With numerical experiments \citet{steiner+al07} have explored the feasibility of 
using high frequency waves for probing the magnetic fields in the photosphere 
and the chromosphere of the Sun. They track an artificially excited, plane-parallel,
monochromatic wave that propagates through a non-stationary, realistic 
atmosphere, from the convection-zone through the photosphere into the 
magnetically dominated chromosphere, where it gets refracted and reflected.

When comparing the wave travel time between two 
fixed geometrical height levels in the atmosphere (representing the formation
height of two spectral lines) with the topography of the surface of equal magnetic 
and thermal energy density (the magnetic canopy or $\beta=1$ surface) we find good 
correspondence between the two. These numerical experiments support
expectations by \citet{finsterle+al04} that high frequency waves bear 
information on the topography of the `magnetic canopy'.
This simulation exemplifies how a piece of Sun can be made accessible
to virtual experimenting by means of realistic numerical simulation.


%
\def\aj{AJ}%
\def\Astronomische Nachrichten Supplement{Astron. Nachr. Suppl.}%
\def\araa{ARA\&A}%
\def\apj{ApJ}%
\def\apjl{ApJ}%
\def\apjs{ApJS}%
\def\ao{Appl.~Opt.}%
\def\apss{Ap\&SS}%
\def\aap{A\&A}%
\def\aapr{A\&A~Rev.}%
\def\aaps{A\&AS}%
\def\azh{AZh}%
\def\baas{BAAS}%
\def\jrasc{JRASC}%
\def\memras{MmRAS}%
\def\mnras{MNRAS}%
\def\pra{Phys.~Rev.~A}%
\def\prb{Phys.~Rev.~B}%
\def\prc{Phys.~Rev.~C}%
\def\prd{Phys.~Rev.~D}%
\def\pre{Phys.~Rev.~E}%
\def\prl{Phys.~Rev.~Lett.}%
\def\pasp{PASP}%
\def\pasj{PASJ}%
\def\qjras{QJRAS}%
\def\skytel{S\&T}%
\def\solphys{Solar~Phys.}%
\def\sovast{Soviet~Ast.}%
\def\ssr{Space~Sci.~Rev.}%
\def\zap{ZAp}%
\def\nat{Nature}%
\def\iaucirc{IAU~Circ.}%
\def\aplett{Astrophys.~Lett.}%
\def\apspr{Astrophys.~Space~Phys.~Res.}%
\def\bain{Bull.~Astron.~Inst.~Netherlands}%
\def\fcp{Fund.~Cosmic~Phys.}%
\def\gca{Geochim.~Cosmochim.~Acta}%
\def\grl{Geophys.~Res.~Lett.}%
\def\jcp{J.~Chem.~Phys.}%
\def\jgr{J.~Geophys.~Res.}%
\def\jqsrt{J.~Quant.~Spec.~Radiat.~Transf.}%
\def\memsai{Mem.~Soc.~Astron.~Italiana}%
\def\nphysa{Nucl.~Phys.~A}%
\def\physrep{Phys.~Rep.}%
\def\physscr{Phys.~Scr}%
\def\planss{Planet.~Space~Sci.}%
\def\procspie{Proc.~SPIE}%
\let\astap=\aap
\let\apjlett=\apjl
\let\apjsupp=\apjs
\let\applopt=\ao


\bibliographystyle{aa}
\bibliography{steiner}

\end{document}